\documentclass[12pt]{article}
\usepackage{amssymb,amsmath,esint,titlesec}
\titleformat*{\section}{\large\bf}
\titleformat*{\subsection}{\normalsize\bf}

\begin{document}


\begin{titlepage}

\setlength{\baselineskip}{24pt}

                               \vspace*{0mm}

                             \begin{center}

{\LARGE\bf  An entropy for groups of intermediate growth}

                            \vspace*{3.5cm}

              \large\sf    NIKOLAOS  \  KALOGEROPOULOS $^\dagger$\\

                            \vspace{0.2cm}
                            
 \normalsize\sf   Center for Research and Applications \\
                                     of Nonlinear Systems \  (CRANS)\\
                          University of Patras, Patras 26500, Greece\\

                            \end{center}

                            \vspace{3.5cm}

                     \centerline{\normalsize\bf Abstract}
                     
                           \vspace{3mm}
                     
\normalsize\rm\setlength{\baselineskip}{18pt} 

One of the few accepted dynamical foundations of non-additive (``non-extensive") statistical mechanics  
is that the choice of the appropriate entropy functional describing a system with many degrees of freedom 
should reflect the rate of growth of its configuration or phase space volume. 
We present an example of a group, as a metric space, that may be used as the phase space of a system 
whose ergodic behavior is statistically described by the recently proposed 
$\delta$-entropy. This entropy is a one-parameter variation of the Boltzmann/Gibbs/Shannon functional
and is quite different, in form, from the power-law entropies that have been recently studied.  
We use the first Grigorchuk group for our purposes.    
We comment on the connections of the above construction with the conjectured 
evolution of the underlying system in phase space.  
         
                           \vfill

               \hspace*{22mm}  Groups of intermediate growth, Self-similar groups, Fractals.    \\
                                                                         
                             \vfill

\noindent\rule{13cm}{0.2mm}\\  
   \noindent $^\dagger$ \small\rm Electronic address: \ \  {\normalsize\sf   nikos.physikos@gmail.com}\\

\end{titlepage}
 

                                                                                \newpage                 

\rm\normalsize
\setlength{\baselineskip}{18pt}

                                              \section{Introduction.}

The idea of entropies that have a functional form which is different from that of the Boltzmann/Gibbs/Shannon (BGS) 
\begin{equation}
                \mathcal{S}_{BGS} [\{p_i \} ] \ = \  k_B \sum_{i\in I} \ p_i \log \frac{1}{p_i} 
\end{equation}
where \ $k_B$ \ stands for the Boltzmann constant and the set \ $I$ \  is used to label the different states, 
has received some attention during the last three decades. 
This is attention is due, in no small part, to the proposal and development for the purposes of Statistical 
Mechanics, and not only, of numerous power-law functionals, the best-known of which is the 
Havrda-Charv\'{a}t \cite{HC} / Vajda \cite{Vaj} / Dar\'{o}czy \cite{Dar} / Lindhard-Nielsen \cite{LN} / Cressie-Read \cite{CR, RC} /Tsallis \cite{Ts, Ts-book}, 
henceforth to be called for brevity just ``Tsallis", entropy. This is given by
\begin{equation}
         \mathcal{S}_q [\{ p_j \} ] \ = \ -k_B \ \frac{1}{1-q} \left( 1- \sum_{i\in I} p_j^q \right) 
\end{equation} 
where \ $q\in\mathbb{R}$ \ or even \ $q\in\mathbb{C}$ \ is called non-extensive or entropic parameter.\\
  
One of the basic problems that all the non-BGS functionals face, is their dynamical foundations. 
Conjectures abound about their dynamical origin and scope of applicability, but little is actually known, 
especially if someone does not perform any numerical experiments/analysis, in this direction \cite{Ts-book}. \\

Let  $| I |$ be the cardinality of the index set $I$, namely the total number of effective states/degrees of freedom 
in the system under study. One of the few model-independent statements that seem to be reasonably supported in this direction 
\cite{Ts-book, HT, TC}    is that the BGS  
entropy is appropriate for systems whose phase space volume increases exponentially, at least in the 
(naive/straightforward) thermodynamic limit, namely
\begin{equation}
         |I(N)| \ \sim  \ A \xi^N, \ \ \ \  \ \ \  A>0, \ \ \xi>1, \ \ N\rightarrow\infty
\end{equation}
following the notation of \cite{TC}. By contrast, power-law entropies should be used for systems whose phase space volume 
increases in a power-law manner
\begin{equation}
    |I(N)| \ \sim \ BN^\tau, \ \ \ \ \ \ \ \   B>0,\ \  \tau>0, \ \ N\rightarrow \infty
\end{equation}
 By hindsight, and if one demands extensivity of the entropic functional $\mathcal{S}$ whatever its exact functional form may be for 
 describing such a system, then 
\begin{equation}
   \mathcal{S}  \ \sim \ N
\end{equation}
 the above two statements are probably trivially true.   \\      
   
Not too long ago, a proposal was made \cite{Ts-book, TC} that uses a modification/generalization of the BGS and Tsallis entropies to capture the 
holographic character of the entropy of black holes. This is a two-parameter functional, that is  pertinent for our purposes, one parameter reduction is  given by     
\begin{equation}
    \mathcal{S}_\delta [\{p_i \}] \ = \ k_B \sum_{i\in I} \ p_i \left( \log \frac{1}{p_i}  \right)^\delta, \ \ \ \ \delta > 0 
\end{equation}
We call the functional (6) the $\delta$-entropy, for brevity, in the sequel. Obviously 
\begin{equation} 
      \lim_{\delta \rightarrow 1} \  \mathcal{S}_\delta [\{ p_i \} ] \  = \ \mathcal{S}_{BGS} [\{ p_i \} ]  
\end{equation}
The $\delta$-entropy has many desirable properties generalizing those of $\mathcal{S}_{BGS}$ and $\mathcal{S}_q$ \cite{Ts-book, TC}. \ 
Irrespective of the applicability of (6) in the case of black hole entropy which it was partially constructed to address, 
the latter's origins being a topic of considerable speculation 
over the last four decades, someone may wonder on what should the growth rate of the set of states / phase space $|I|$ 
of a system be, so that it is described by the entropic functional (6). An answer is provided in \cite{Ts-book, TC}:
\begin{equation}
        | I(N) | \ \sim \ \phi(N) \nu^{N^\gamma}, \ \ \ \ \ \ \     \nu>1, \ \ 0 < \gamma < 1
\end{equation}
with $\phi(N)$ chosen such that 
\begin{equation}
      \lim_{N\rightarrow\infty} \frac{\log \phi (N)}{N^\gamma} \ = \ 0 
\end{equation}
and
\begin{equation}
     \delta \ = \ 1/\gamma
\end{equation}
The question that naturally arises is: are there systems whose phase space / set of states grow, at least asymptotically \ (as \ $N\rightarrow\infty$), \
by a law given by (8)?  This law describes growth rates which are slower than exponential (3) but faster than any 
power-law (4). Do such systems even exist, or the growth law (8) is too artificial to be of any physical or mathematical interest?\\

We have attempted to make the present work somewhat self-contained and readable our intended audience of Physicists. In Section 2, we present some 
general comments about the dynamics of phase spaces and the difficulties associated with such calculations. In Section 3, we present background, 
the construction and aspects of the groups of intermediate growth. Sections 4, contains some general comments and an outlook toward future work. \\   


\section{Dynamics on the phase space manifold.}

One would, ideally, like a dynamically derived answer to the above question. Namely, pick a conventional phase space (which is an even-dimensional 
symplectic manifold), or a set of states in the Hilbert or Fock space of a quantum system, and determine the evolution, or probability 
distribution, that would give rise to (8). In both cases the number of degrees of freedom should allow \ $N\rightarrow\infty$ \ 
as the naive thermodynamic limit.   
Such a calculation is practically intractable even for the simpler systems 
that are described by $\mathcal{S}_{BGS}$  where ad hoc assumptions, like the ergodic hypothesis, have to be invoked to allow the calculation to 
proceed. Given this experience, any such attempt toward computation of the  power-law (4) or sub-exponential growth rates would be a doomed 
exercise in dynamics, at our current level of understanding and available techniques for such systems.\\
 
To proceed we modify the question and ask the following: assume that we are given a dynamical system which is ergodic. What properties should 
its phase space have to allow for a growth rate given by (8)? This modified question reduces a dynamical into a kinematical question: a question about he dynamical 
evolution of the system in phase space essentially reduces to a geometric question about the growth rate of the system's volume. As a result, the initially intractable problem 
is highly simplified and may be more amenable to analysis. 
The issue of the physical relevance of such a modified question should not be forgotten in  this modified context. \\         

If we use a classical rather than a quantum system, for simplicity, the author knows of no results in the category of even-dimensional symplectic manifolds 
of high dimension that would allow for the sub-exponential  increase in the phase space volume. We recall \cite{Arn, HZ} that a symplectic $n-$dimensional manifold 
\ $(\mathbf{M}, \omega)$ \ has a natural volume form: the Liouville measure \ $\omega^n$. \ However a symplectic manifold does not possess any 
local invariants under symplectic diffeomorphisms since it is locally equivalent to \ $\mathbb{R}^n$, \ which is the content of Darboux's theorem. 
As a result, a Hamiltonian flow  \ $\varphi_t: \mathbf{M} \times [0,1] \rightarrow \mathbf{M}$ \  preserves the symplectic volume 
(Liouville's theorem): \ $\varphi_t (\omega) = \omega$. \  Analyzing the volume growth for symplectic manifolds in order to get an idea for 
which entropic form, or class of entropic forms, needs results on the global aspects of symplectic manifolds \cite{HZ} and 
may therefore be of questionable utility toward our goal. \\ 
  
Alternatively, one can take Riemannian/metric viewpoint (see, for instance \cite{Pettini} and references therein) 
and attempt to analyze the phase space seen as a Riemannian manifold $(\mathbf{M, g})$. 
 The obvious drawback in this approach is that, generically,  there is no natural phase space metric \ $\mathbf{g}$. \ Introducing such a \ $\mathbf{g}$ \ 
 is done in an ad hoc manner despite the fact that its presence determines the volume growth rate in a very essential way. 
 This drawback is somewhat mitigated by the  fact that  many Hamiltonians of physical interest involve a quadratic kinetic term which is can be 
 used to provide the required Riemannian metric in this approach in a more natural (although still non-unique) way.\\
  
 The advantage of the metric approach is that Riemannian manifolds have local structure: the sectional curvature, or equivalently the Riemann tensor,
 quantifies locally the deviation of the manifold from flatness. 
 To proceed, one may to directly solve the differential equation (geodesic equation, in the simplest cases)  
 describing the evolution of the system if the underlying manifold is simple enough. 
 Alternatively, one can use comparisons with appropriately chosen  simply connected manifolds of constant sectional curvature 
 (space forms) to attain bounds on the growth rate of the volume of the phase space.
 The former method is much more desirable, but practically impossible to explicitly implement in almost any case of physical interest.
 Hence someone may use extremely  simplified models of phase spaces, such as symmetric or homogeneous spaces, whose additional 
 group theoretical structure makes them amenable to further analysis \cite{Hel}. \\
 
 We have come to realize by using comparison methods \cite{CE, GP, Gr-book},  that there is a sharp distinction between manifolds of positive 
 and manifolds of negative
 sectional curvature. The volume grows in a power-law manner in Euclidean (flat) spaces but exponentially for manifolds of 
 negative curvature \cite{CE, GP, Gr-book} as can be readily seen either by explicitly solving the Jacobi  equation, or by using the Bishop-Gromov 
 volume comparison theorem. However most 
 instances of physical interest involve manifolds having curvature of variable sign and without an obvious upper or lower bound in their sectional or Ricci
 curvatures \cite{Pettini}. Such lack of sign definiteness and explicit bounds  makes the application of the comparison techniques ineffective, 
 at least directly.  It is also suspected that it is the  variation 
 of the sign of the sectional curvature that is crucial for many of the dynamical properties of the underlying system \cite{Pettini}. 
 In this spirit, a delicately chosen succession of 
 regions of positive and negative curvature in a manifold, possibly resulting from geometric surgery, may be able to produce the sought after volume growth rate (8).  
 However the author is not aware of any general results or any examples of manifolds in which such volume growth (8) has even be conjectured, let alone 
 demonstrated or explicitly calculated.\\     

 If someone insists on having a metric and a symplectic structure which are compatible with each other on the phase space, and also integrable,
 then one is naturally lead to demand the phase space to be a K\"{a}hler manifold \cite{Wells, Ball}. 
 The existence and compatibility between the complex, the metric and the symplectic structures introduces a substantial rigidity in the set of K\"{a}hler 
 manifolds. Due to such rigidity, we do not naively expect that K\"{a}hler manifolds can provide options for the volume growth that the Riemannian ones  do not already 
 exhibit.\\     

As a result, one may have to broaden the class of spaces that should be considered as phase spaces of ergodic systems described by (6).  One class of such 
spaces are general metric measure spaces \cite{Gr-book}, but this class also seems to be too broad and constructions are far less rigid and difficult  to control. 
Pragmatically, someone can also question the usefulness of general metric measure spaces above and beyond that of symplectic and / or Riemannian manifolds 
for physical purposes, even though such spaces may be important in the thermodynamic limit.   \\ 


\section{The $\delta$-entropy and groups of intermediate growth.} 

A class of objects that have been extensively studied in Mathematics over the last two centuries and have  also been extensively used in Physics, are groups. 
One would be hard pressed to find a field of Physics in which groups and group theoretic arguments play a minor role. In this work we will be interested
in discrete rather than continuous (topological, Lie etc) groups. It turns out that such (usually infinite) groups, which have a finite number of generators 
have strong geometric properties that have been explored for more than an century.  In particular, one can use geometric constructions and the concomitant 
arguments based on these groups or acting on metric and measure spaces  to infer or elucidate some of their algebraic aspects. 
This is the goal of combinatorial \cite{MCS, LS} and subsequently geometric group theory \cite{delaH}
which is a subject with substantial history and successes. \\

During the last three decades there has been a revival  of the activity on the  geometric properties of such groups or their geometric actions 
on metric measure spaces, due in no small part to the influence of \cite{Gr1, Gr2}.  It turned out that some finitely generated groups (Grigorchuk groups) 
have the flexibility, and at the same time the rigidity, required to provide explicit examples that fit our purposes. 
The background and all results in the rest of the paper follow \cite{GrigPak, Grig2013} 
which are very readable expositions  and the references therein; some developments that may be somehow relevant to our goals 
which are  related to groups of intermediate growth can also be  found in the expository \cite{Nekr-book, Nekr-Tepl, Kaiman, Grigor-Dyn, BGN} 
among the numerous references in the field. \\


\subsection{Preliminaries}

 Consider a finitely generated group $\mathbb{G}$ and a symmetric generating set  $S=\{s_1, s_2, \ldots, s_n \}$. 
 The group $\mathbb{G}$ is assumed to be finitely generated,  but not necessarily finitely presented. 
 Actually, as will be stated later, there is no known finite presentation of the Grigorchuk group. By definition, any element  \ $g\in \mathbb{G}$ \ 
 can be expressed, non-uniquely,  as a product of elements of $S$
 \begin{equation}         
         g \ = \ s_{i_1} s_{i_2} \cdots s_{i_n}   
 \end{equation}
The (word) norm $\| g \|$ of \ $g\in\mathbb{G}$ \ with respect to the generating set \ $S$, \ is defined to be the length of the shortest word presentation of \ $g$ \ 
in terms of elements of \ $S$. \  Equivalently,  this is the combinatorial distance from the identity element of \ $\mathbb{G}$ \ to \ $g\in\mathbb{G}$. \
Clearly the word norm of \ $g\in\mathbb{G}$ \ depends on the generating set \ $S$. \  However, it turns out that picking a different generating set 
changes the norm in a ``controllable" way; for two different generating sets \ $S_1$, \ $S_2$ \ with corresponding norms $\| g \|_1, \| g \|_2$ one gets
\begin{equation}         
       \| g \|_1 \ \leq \ C \| g \|_2, \ \ \ \ \ \ \ \| g \|_2 \ \leq \  C \| g \|_1
\end{equation}
for some constant \ $C>0$. \ Such relations are not uncommon in several branches of Mathematics. In a previous work \cite{NK1} we commented on a similar relation 
holding between any two norms of a linear space, its physical motivation, and its implications in showing that the metric of space-time has to obey the 
Pyrthagorean theorem. The word metric on \ $\mathbb{G}$ \  with respect the generating set \ $S$ \  is given by 
\begin{equation}     
      d_S (g, h) \ = \ \| g^{-1}h \|_S, \ \ \ \  \ \ \ \ \forall \ g, h \in \mathbb{G}  
\end{equation}

Let \ $\Gamma(\mathbb{G}, S)$ \ be the Cayley graph of \ $\mathbb{G}$ \ with respect to the generating set \ $S$: \ this is the graph whose vertex set \ $V$ \ 
is the set of elements of \ $\mathbb{G}$, \ where an edge of \ $\Gamma (\mathbb{G}, S)$ \ is drawn between any two elements of \ $\mathbb{G}$ \ that are 
connected by an element of \ $S$. \ 
The distance between any two elements of \ $\Gamma(\mathbb{G}, S)$ \  is the shortest path between these two elements, if each edge of the Cayley graph 
is assumed to have a length one unit. If one considers \ $\mathbb{G}$ \ with two different generating sets, the two Cayley graphs may be quite different as metric 
spaces: there is no reason, for instance, for them to be homeomorphic, with the assumed open-set topology, let alone isometric.  However, one can prove, based 
on the fact that  \ $\mathbb{G}$ \ is finitely generated, that two such Cayley graphs are quasi-isometric \cite{delaH}. 
Quasi-isometries are large scale maps of controlled distance distortion. 
Let \  $f: \mathfrak{X} \rightarrow \mathfrak{Y}$ \ be a (not necessarily continuous) map, where \ $\mathfrak{X, Y}$ \ are metric spaces with distance functions \ 
$d_\mathfrak{X}, d_\mathfrak{Y}$ \  respectively. Then \ $f$ \ is a quasi-isometric embedding if there are constants \ $C_1 > 1, \  C_2 >0$ \ such that       
\begin{equation}
   \frac{1}{C_1} d_\mathfrak{X} (x,y) - C_2 \ \leq \ d_\mathfrak{Y} (f(x), f(y)) \ \leq \ C_1 d_\mathfrak{X}(x,y) + C_2, \ \ \ \ \ \ \forall \ x, y \in \mathfrak{X}
\end{equation}
If, moreover, \ $f$ \ is almost surjective, namely if there is a constant \ $C_3 > 0$ \ such that  
\begin{equation} 
      d_\mathfrak{Y} (z, f(x)) \ \leq \ C_3, \ \ \ \ \ \ \forall \ x\in\mathfrak{X}, \ z\in\mathfrak{Y}
\end{equation}
then \ $f$ \ is called a quasi-isometry. One can see that \ $\Gamma (\mathbb{G}, S)$ \ is regular and that \ $\mathbb{G}$ \ acts transitively on it, hence 
\ $\Gamma(\mathbb{G}, S)$ \ is  homogeneous.\\


\subsection{About growth rates} 

Having defined  a distance function \ $d_S$ \ on \ $(\mathbb{G}, S)$, \ one can define the ball with center \ $g\in\mathbb{G}$, \ indicated as 
\ $B_g(n) \subset \mathbb{G}$, \ of radius \ $n\in\mathbb{N}$ \ to be 
\begin{equation}   
           B_g(n) \ = \  \{  h \in \mathbb{G}: \ d_S (g,h) \ \leq \ n   \}
\end{equation}
Since $\mathbb{G}$ acts transitively on itself by (left) multiplication, we need to consider only balls that are centered at the identity of $\mathbb{G}$, 
 indicated by  \ $B(n)$. \    The cardinality (counting measure) of \ $B(n)$ \ provides a natural measure on  the ball.  Then,  the growth function of 
 \ $(\mathbb{G}, S)$ \  is defined by  
\begin{equation}
        \gamma_{(\mathbb{G}, S)} (n) \ = \ \mathrm{card} \ B(n)
\end{equation}
Using (12), one can see that the growth functions for two different generating sets \ $S_1, \ S_2$ \ of \ $\mathbb{G}$ \ are related, with $C>0$,  by 
\begin{equation}
    \gamma_{(\mathbb{G}, S_1)} (n) \  \leq \ \gamma_{(\mathbb{G}, S_2)} (Cn), \ \ \  \ \ \  \gamma_{(\mathbb{G}, S_2)} (n) \  \leq \ \gamma_{(\mathbb{G}, S_1)} (Cn) 
\end{equation}
To eliminate the continuous repetition of such bounds, one can use an equivalence better adapted to (12), (18) than the standard equality. Two functions   
$\gamma_i: \mathbb{N} \rightarrow \mathbb{N}, \ i=1,2$ are equivalent, indicated by \ $\gamma_1 (n) \sim \gamma_2 (n),$ \ if there is a constant \ 
$c \in \mathbb{N}$ \ such that  
\begin{equation}
    \gamma_1(n)  \ \leq \gamma_2 (cn), \ \ \ \ \ \gamma_2(n) \ \leq \gamma_1(cn)
\end{equation}
In this terminology, the growth functions of two generating sets \ $S_1$ \ and \ $S_2$ \ of \ $\mathbb{G}$ \ are equivalent, so if we are interested in 
such growth rates up to equivalence, we may skip the explicit mention of the generating set and write \ $\gamma_\mathbb{G}$ \ instead of 
 \  $\gamma_{(\mathbb{G},S)}$ \ which we will be doing in the sequel.        
Since any finitely generated group can be seen as being the quotient of the free group of $k$ generators by relations, 
then the highest possible growth rate of any such \ $\mathbb{G}$ \ is that of the free group, namely 
\begin{equation}   
     \gamma_\mathbb{G} (n) \ \leq \ \sum_{i=0}^n\  (2k)^i \ \leq \ (2k+1)^n   
\end{equation}
A function \ $f: \mathbb{N} \rightarrow \mathbb{N}$ \ is  polynomial if 
\begin{equation} 
      f(n) \sim n^\alpha, \ \ \ \ \ \ \alpha > 0
\end{equation}
An example of such a case is a polynomial function \ $f(x) = a_m x^m + a_{m-1} x^{m-1} + \ldots + a_1x + a_0$, \ with \ $m\in\mathbb{N}$. \ 
Then such \ $f$ \ obeys \ $f(x) \sim x^m$. \ A function \ $f: \mathbb{N} \rightarrow \mathbb{N}$ \ is called super-polynomial if  
\begin{equation}
      \lim_{n\rightarrow\infty} \frac{\log f(n)}{\log n} \ = \ \infty
\end{equation}
Examples of super-polynomial functions are \ $f(n) = n^n$, \ and \ $f(n) = n^{\log\log n}$. \  A function \ $f: \mathbb{N} \rightarrow \mathbb{N}$ \
is called exponential, if 
\begin{equation}  
    f(n) \ \sim \ e^n
\end{equation}
An example of such a function is \ $f(x) = x^n e^x$. \ A function \ $f: \mathbb{N} \rightarrow \mathbb{N}$ \  is called sub-exponential if 
\begin{equation}
    \lim_{n\rightarrow \infty} \frac{\log n}{n} \ = \ 0 
\end{equation}
 An example of such a function is \ $f(x) = e^\frac{\log n}{n}$. \  It may be worth noticing that there are functions that do not fall in any of the above 
 classes such as \ $e^{n^{\sin n }}$, \ for instance, as this oscillates between \ $1$  \ (which is polynomial)  and \ $e^n$ \ (which is exponential). 
 A function \ $f: \mathbb{N} \rightarrow \mathbb{N}$ \ has intermediate growth if it is sub-exponential and super-polynomial at the same time. An 
 example of such a function is \ $f(x) = e^{\sqrt{x}}.$  So, the statement of (20) can be re-expressed by stating that any group has at most an 
 exponential growth rate.    \\   

By the definitions of the above paragraph one can classify the growth rate of groups into polynomial, intermediate and exponential.  
We can immediately see that the growth of a group $\mathbb{G}$  is the same as the growth of a subgroup of finite index. 
Moreover, the growth of $\mathbb{G}$ is not smaller than the growth of a finitely generated subgroup. Some comments regarding
 growth rates are
\begin{itemize}   
    \item[$\circ$] The power-law functions \ \ $f(n) \sim n^\alpha$ \ \ belong to different equivalence classes for different values of \ $\alpha \geq 0$.  
    \item[$\circ$] All exponential functions \  \ $f(n) \sim x^n, \ \ x>1$ \ \ are equivalent and belong, therefore, to the same class.  
    \item[$\circ$] All functions of intermediate type  \ \ $e^{n^\alpha}, \ \ 0< \alpha <1$ \ \ belong to different equivalence classes. 
\end{itemize}


\subsection{Group growth, extensivity and entropies}

One can look at the above equivalences from the viewpoint of entropies regarding the proposal of \cite{HT} in combination with the results and 
conclusions of \cite{NK2, NK3, NK4, NK5, NK6, NK7, NK8}. The main point is that a functional $\mathcal{S}$ that can legitimately be called ``entropy" 
has to be extensive, namely 
 \begin{equation}    
        \mathcal{S} \sim N, \ \ \ \ \  N\rightarrow \infty
\end{equation}
where \ $N$ \ stands for the number of degrees of freedom of the system. This requirement is  weaker than (5) hence more generally applicable. 
The above, (25), is a requirement of asymptotic extensivity which is 
a reasonable requirement  so that the system has a good macroscopic behavior.  Since (25) should be valid, we can see that 
systems whose growth rate is exponential, assuming also the validity of the ergodic hyppothesis, 
have to be described by an entropic functional which is linear in the logarithm, such as in the BGS entropy case (1). 
By contrast, systems whose growth rate is power-law, and which are assumed to  be ergodic in a subspace of the full phase space,   
due to non-trivial interactions or correlations or any other reason, have to be described by a  power law functional such as (2)
in order to ensure the validity of the asymptotic extensivity condition (25). As we argued in  \cite{NK2, NK3}, the adoption of the Tsallis entropy
can be seen either as a non-trivial addition $+$ of independent systems, or more ``covariantly" as a generalized addition $\oplus_q$. 
Imposing the distributivity requirement between addition and multiplication leads someone to define a generalized product $\otimes_q$ \  which alongside 
the addition $\oplus_q$ on the set $\mathbb{R}$ form a field \  $\mathcal{R}_q = (\mathbb{R}, \oplus_q, \otimes_q)$. The isomorphism between 
$\tau_q: \mathbb{R} \rightarrow \mathcal{R}_q$ is given by 
\begin{equation}
    \tau_q(x) \ = \ \frac{(2-q)^x - 1}{1-q}
\end{equation}        
The effect of this exponential map  is  the following: it shrinks the distances in \ $\mathcal{R}_q$ \ by a logarithmic factor. So, the systems 
that are described by (1) whose volume growth is exponential do get  a volume growth which becomes power-law. In some sense this is the origin 
of the power-law entropies: we  start with a hyperbolic dynamical system. Then due to the dynamics of the underlying system and for reasons 
that are not well-understood, we are forced to consider the volume growth under the inverse of (26) thus getting a system with power-law volume 
growth which is described by (2). The analogy with the group growth may not be lost at this point: all finitely generated groups can be seen as quotients 
of the free group with a number of generators. If the resulting group is virtually nilpotent then it has a power-law volume growth, and any (discrete) 
dynamical system which is ergodic on such a phase space will similarly have a power-law growth. Then its statistical properties will be described 
by a power-law entropy like (2). The novel part  is that from the viewpoint of group volume growth all exponential growth rates are equivalent. But the 
power-law rates are not. As a result \ $\tau^{-1}_q$ \ which gives rise to an entropic functional such as (2) is probably able to distinguish finer 
features of a system, as opposed to (1). This is reflected in the multitude (the ``q-triplet" \cite{Ts-book, Ts-tripl}) of non-extensive parameters each of which can describe 
different aspects of a system.  This was partly our motivation for conjecturing that a change of \ $q$ \ should be a sufficient indication for a phase transition \cite{NK7}. \\

It is also worth seeing that something similar is true for groups of intermediate growth: the mapping from an exponential to an intermediate volume growth rate 
which is induced by an entropic functional such as (6), seems to be able to uncover a more refined structure of the underlying dynamics than (1) used 
in the exponential growth rate case. This because  functions of intermediate growth of different indices \ $\alpha$ \ are inequivalent. To reverse the argument, we claim the following:
the entropies that are used in group theory have always the BGS form (1). However, if one wishes to to have a chance establish, non-trivial results \cite{Grig2013}               
then an entropy of the form (6), or of a functional form equivalent to it,  may have to be employed. This should be done in in order to always maintain (25) 
which seems to be the most robust and absolutely indispensable condition on the properties  of the proposed entropic functionals. \\ 


\subsection{The (first) Grigorchuk group} 

We proceed by  stating the construction of the first Grigorchuk group \cite{GrigPak, Grig2013}, not in the form in which it was initially proposed, 
but in the one widely employed which relies on automorphisms of binary trees. We attempt to provide a possible physical interpretation/analogy 
wherever possible.  We have to state from the outset that there are uncountably many groups of intermediate growth, that the growth rates of most of them 
are not exactly known and that most of such constructions rely on actions (automorphisms) of homogeneous rooted trees.  For the case of the (first) 
Grigorchuk group \  $\mathcal{G}$ \  consider the set of two elements, indicated by $X = \{0, 1\}$. Then consider the set of all possible words $X^\ast$ 
on $X$, namely the set       
\begin{equation}
    X^\ast \  = \ a_1a_2  \ldots a_n \ldots \ \ \ \ \ \ \ a_i \in X, \ \ \ \ i\in \mathbb{N}\cup\{ 0 \}  
\end{equation}
This can be seen as a binary tree \ $T$ \ with the root being the empty set. The set of vertices of \ $T$ \ is the set of the elements of \ $X^\ast$ \ and 
an edge is drawn between two vertices of \ $T$ \  if the vertices have the form \ $a_0\ldots a_n$ \ and $a_0\ldots a_nx$ where \  $x\in X$.  \ Assume that 
each edge has unit length. Then the length of the word \ $L(a_0\ldots a_n)$ \  is equal to \ $n$, \ which indicates how far away it is from the root in the 
length metric. All elements of \ $T$ \ having the same length \ $n$ \ belong to the same level as they are equidistant from the root. 
Let \  $T_v$ \ indicate the sub-tree of \ $T$ \  whose root is the element \ $v\in T$ \ instead of the root. The whole tree is indicated by 
\  $T_{\emptyset}$ \  in this notation, the first level subtrees by \ $T_0$ \ and \ $T_1$ \ etc.  
Due to the fact that \ $T$ \ is infinite, one can see (intuitively obvious) that \ $T_v, v\in T$ \ is isomorphic to \ $T$. \ An automorphism of \ $T$ \ is a bijective map 
that fixes the root and preserves the level of each vertex. Hence such an automorphism is a permutation of subtrees that are rooted at the same level, so
it is pictorially a ``twist" of the branches of the tree, guaranteeing that adjacent vertices remain adjacent.  Let the set of automorphisms of \ $T$ \ be indicated by 
\ $Aut(T)$. \ This has the cardinality (power) of the continuum. It is clear that the action of an element of \ $Aut(T)$ \  at the vertices of a given 
level also depends on its action on all the previous levels. So, very much like the tree \ $T$,  \ the set of automorphisms which can be seen to form a group, 
maintains in some sense ``infinite memory", namely the path that was followed since it was created from the root.  
Therefore, the above, it may be an appropriate construction for processes that have memory, some of which  (2) conjecturally describes. \\

To go straight to the 
heart of the matter,  the first Grigorchuk group is the finitely generated infinite group \ $\mathcal{G}$ \ which is generated, recursively, by four automorphisms indicated by  \ 
$a, b, c, d$ \ which act on \ $a_ix \in X^\ast$ \ with \ $a_i = 0, 1$ \ and \ $x\in X^\ast$ \ or on \ $T$ \ as follows:
\begin{equation}   
             \begin{array}{ll}     
    a(0x) \ = \ 1x,                 &      a(1x) \ = \ 0x \\
    b(0x) \ = \ 0a(x),   \ \ \     &      b(1x) \ = \ 1c(x)\\
    c(0x) \ = \ 0a(x),              &     c(1x) \ = \ 1d(x)\\
    d(0x) \ = \ 0x,                  &     d(1x) \ = \ 1b(x) 
             \end{array} 
\end{equation}  
If one considers the action of each of these automorphisms on the left and right trees having the same root, they can also be rewritten as
\begin{equation}
      b \ = \ (a,c), \hspace{15mm}c \ = \ (a, d), \hspace{15mm} d = (I, b) 
\end{equation} 
where $I$ stands for the identity map. 
We see that $a$ interchanges/switches two subtrees of $T$ of the same level having the same root and that $d$ acts as the identity 
on all words starting with $0$. As an example of the action of these automorphisms, consider how they act on an element of $T$ such as $(1101)$. 
We have:
\begin{displaymath} 
   a(1101)=0101, \ \ \ \ b(1101)=1c(101)=11d(01)=1101 
\end{displaymath}
\begin{displaymath}
c(1101)=1d(101)=11b(01)=110a(1)=1100, \ d(1101) = 1b(101) = 11c(01) = 110a(1) = 1100
\end{displaymath}
One can prove that the automorphisms (28) obey the  relations
\begin{equation}
     a^2 = b^2 = c^2 = d^2 = I
\end{equation}
and 
\begin{equation}
     cd = dc = b, \ \ \ \ \ db = bd = c, \ \ \ \ \ bc = cb =  d 
\end{equation}
Moreover
\begin{equation}
     bcd \ = \ I
\end{equation} 
 and the group generated by \ $b, c, d$ \ is \ $\mathbb{Z}_2^2$. \ Furthermore 
\begin{equation}
   (ad)^4 \ = \ (ac)^8 \ = \ (ab)^{16} \ = \ I
\end{equation} 
Numerous other properties are known about \ $\mathcal{G}$ \cite{Grig2013}, the most important of which, for our purposes, is that it has an intermediate 
growth rate, a fact which is not trivial to prove \cite{GrigPak}.   \\


\subsection{A ``physical" interpretation and further comments}

We  try to provide a more ``physical interpretation" of some of the constructions above. First of all, the fact that we use an infinite  tree
can be interpreted as indicating that the system under study has infinite memory. A binary option can be interpreted as a fermionic state which is 
either empty $(0)$ or full $(1)$. We use here the word ``fermionic"  as a pure analogy, since no spacetime symmetries exist anywhere in our considerations. 
Any word of the alphabet $X^\ast$ of any vertex of the tree $T$ can be seen as a fermionic multi-particle state, namely an
element of some   Fock space. The difference between this construction and the more conventional one is that here the system has infinite memory. 
Here, we do not have in our disposal creation and annihilation operators which act locally on this Fock space.  Instead, we should keep track of 
all previous steps that lead to a particular configuration, starting for the root (the vacuum). States having the same length can accommodate, at most, 
the same number of fermions. As one moves further away from the root of the tree, then more and more states are available and may be filled or not.  \\
  
The conventional creation and annihilation operators are composite in this description. To fill an empty single fermionic slot, someone would have to move back 
one step toward the root of the tree and then forward one more step along the complementary branch of the subtree. The same applies to annihilating a single
fermionic state in such a multi-particle analogy. This is, in essence, what the automorphism \ $a$ \ of the tree \ $T$ \ accomplishes. 
Due to the binary nature (occupied by zero or one  fermions) of the fermionic states, it is not surprising that \ $a$ \ is an idempotent ($a^2 = 1$). 
Then from (28) one can provide a similar in spirit, but far less clear, interpretation of the automorphisms \ $b, c, d$. \\

 This lack of a familiar interpretation of the generators \ $b, c, d$ \ takes place essentially because the 
Grigorchuk group \ $\mathcal{G}$ \ is not a linear group, namely it is not a subgroup the general linear \ $n\times n$ \ (matrix) group \ $GL_n(F)$, \ where \ $F$ \ 
is a field of zero characteristic. According to the Tits alternative,  a linear, finitely generated group has either polynomial growth if it is virtually solvable, 
or it contains a non-Abelian free subgroup, so it has exponential growth. There is no possibility for such a group to have intermediate growth
as \ $\mathcal{G}$ \ does. Hence, even if  one may initially  suspect that due to (30), (31), (32) the generators \ $b, c, d$ \ may act in a analogous way to 
real generators of an \ $N=4$ \ supersymmetry algebra, or as the set of almost complex structures of a hyper-K\"{a}hler manifold, up to complexification and 
normalization, there will invariably be further properties such as (33) that would immediately spoil such a physically familiar  interpretation.\\

Physics extensively uses ``infinite" groups: 
the gauge groups in Particle Physics, the diffeomorphism group, the Bondi-Metzner-Sachs group etc in General Relativity, the Kac-Moody and loop groups 
in string theories etc are all infinite dimensional. However \ $\mathcal{G}$ \ is still quite different from all of the above in that it is pure torsion: it may be finitely generated 
and infinite, but all its elements have a finite order which is a power of $2$. Moreover \ $\mathcal{G}$ \ is ``just infinite", in the sense that any subgroup of it is finite.
and in addition \ $\mathcal{G}$ \  is itself residually finite. \\

The fact that \  $\mathcal{G}$ \ as well as all other known groups of intermediate growth are infinitely presented does not necessarily exclude their use in Physics, 
due to the existence of the recursion relations (28). Moreover there is substantial overlap between constructions related to groups of intermediate growth and 
self-similar groups \cite{Nekr-book}. Actually \ $\mathcal{G}$ itself is a self-similar group. What is intriguing, and potentially very physically relevant, is that 
such self-similar constructions give rise to fractals, objects of considerable interest especially if one recalls that fractals were one of the motivations that lead 
to the introduction of (2).   \\
                   
It may be of  physical interest to see whether any of the more familiar phase spaces (finite dimensional Riemannian manifolds) can possibly have 
intermediate volume growth. A theorem of A. Avez, responding to a conjecture of E. Hopf, seems to exclude such a possibility, for the compact case at least:  
a compact, connected Riemannian manifold \ $\mathbf{M}$ \ without focal points, as the phase space of a Hamiltonian system should be, has either a fundamental group of 
exponential growth or is flat. The former option implies that the manifold itself, up to quasi-isometry,  has a universal cover which has the same 
exponential growth. In the latter case the manifold has polynomial growth. Hence, a compact manifold of intermediate growth does not exist. However, 
not much seems to be known about the non-compact case, which is pertinent to our goals, and phase spaces of physical systems may very well be non-compact.  
All these comments  assume that the underlying evolution of a physical system is ergodic on its phase space. 
We see it as very likely likely that ``complex systems" will have a non-ergodic evolution in their phase spaces \cite{Ts-book, NK9}, 
so we may have to use  other concepts and techniques of determining their behavior than the ones presented here.\\       


\section{Discussion and speculations}

In this work we argued that an entropic functional ($\delta$-entropy) that was introduced almost a decade ago \cite{Ts-book} and was recently considered in 
relation to the black hole entropy \cite{TC} may be appropriate for the description of dynamical properties of ergodic systems on  spaces of intermediate volume growth.
We presented the Grigorchuk group as the exemplar and best known case for such a space of intermediate growth and commented on the  potential physical 
implications and interpretation that its use may entail. Conversely, we suggested the use of the $\delta$-entropy for probing properties of such spaces of 
intermediate growth.\\     

The present work can be placed in the wider context of examining the dynamical foundation of statistical mechanics, particularly in determining the basis and
range of applicablility of the non-additive entropic forms that have been developed over the last three decades in the Physics community. A common
weakness of these approaches is the complete lack of analytically tractable models of systems with many degrees of freedom that can be studied
as examples  and for obtaining physically relevant results. The present work also suffers from such a lack of concrete examples. On the  other hand,
it expands the range of systems and techniques at which one may usually look  for properties of systems that may be described by such non-additive entropies.  
The connection between groups of intermediate growth, self-similar groups and fractals may be  potentially quite promising in addressing
some of the important issues that have arisen since the introduction of (2) and several of the other non-additive entropies during the last few decades.\\   
  
  
                          \newpage
  
 \noindent{\normalsize\bf Acknowledgement:} \  The author is greatly indebted and grateful  to the Director of CRANS, Professor Anastasios Bountis, for his constant  
 inspiration, encouragement and support as well as for many fruitful conversations without which this work would have never been possible.\\   
  
  
 \noindent{\normalsize\bf Conflict of interest:} \ The author declares that there is no conflict of interest regarding the publication of this paper. \\ 
  



\end{document}